\documentclass[a4paper,conference,final]{IEEEtran}

\usepackage[pdftex]{graphicx}
\usepackage{array}
\usepackage{multirow}
\usepackage{amssymb}
\usepackage{amsmath}
\usepackage{euscript}

\usepackage[draft]{pgf}

\newcolumntype{P}[1]{>{\centering\arraybackslash}p{#1}}
\newcolumntype{M}[1]{>{\centering\arraybackslash}m{#1}}

\begin{document}

\title{Aggregation and Trunking of M2M Traffic \\ via D2D Connections}

\author{\IEEEauthorblockN{Giovanni Rigazzi\IEEEauthorrefmark{1}, Nuno K. Pratas\IEEEauthorrefmark{2}, Petar Popovski\IEEEauthorrefmark{2}, Romano Fantacci\IEEEauthorrefmark{1}}
\IEEEauthorblockA{\IEEEauthorrefmark{1}Department of Information Engineering, University of Florence, Italy, E-mail: name.surname@unifi.it}
\IEEEauthorblockA{\IEEEauthorrefmark{2}Department of Electronic Systems, Aalborg University, Denmark, E-mail: \{nup, petarp\}@es.aau.dk}}

\maketitle

\begin{abstract}

Machine-to-Machine (M2M) communications is one of the key enablers of the Internet of Things (IoT). Billions of devices are expected to be deployed in the next future for novel M2M applications demanding ubiquitous access and global connectivity.
In order to cope with the massive number of machines, there is a need for new techniques to coordinate the access and allocate the resources. Although the majority of the proposed solutions are focused on the adaptation of the traditional cellular networks to the M2M traffic patterns, novel approaches based on the direct communication among nearby devices may represent an effective way to avoid access congestion and cell overload. In this paper, we propose a new strategy inspired by the classical Trunked Radio Systems (TRS), exploiting the Device-to-Device (D2D) connectivity between cellular users and Machine-Type Devices (MTDs). The aggregation of the locally generated packets is performed by a user device, which aggregates the machine-type data, supplements it with its own data and transmits all of them to the Base Station. We observe a fundamental trade-off between latency and the transmit power needed to deliver the aggregate traffic, in a sense that lower latency requires increase in the transmit power. 
\end{abstract}


\section{Introduction} \label{sec.introduction}
Machine-to-Machine (M2M) refers to a novel class of autonomous communications involving multiple devices, which interact among themselves without human intervention. The M2M paradigm opens the door to many applications, ranging from Smart Grid, Health Monitoring to Intelligent Transport System, and fosters the interconnection among diverse systems within heterogeneous networks. This has gradually drawn the attention of academia and industry, as well as of standardization bodies that are currently focusing on air interface improvements towards a full M2M support~\cite{6163599}.

Typically, M2M systems are characterized by a massive number of deployed devices, with different capabilities in terms of computation, battery, and coverage. As a consequence, a huge amount of data is expected to be generated, processed and delivered through different networks. Furthermore, heterogeneous applications imply different requirements in terms of QoS, latency, reliability, and security. For example, health monitoring requires mostly low-latency and highly reliable packet delivery due to the critical information carried and the limited capacity of the devices employed~\cite{4736537}. On the other hand, in a large class of M2M applications, the traffic consists of a high number of packets characterized by a short payload, which compromises the efficiency of traditional systems designed to support Human-to-Human (H2H) traffic.
\begin{figure}[t]
\begin{center}
\includegraphics[width=0.6\columnwidth]{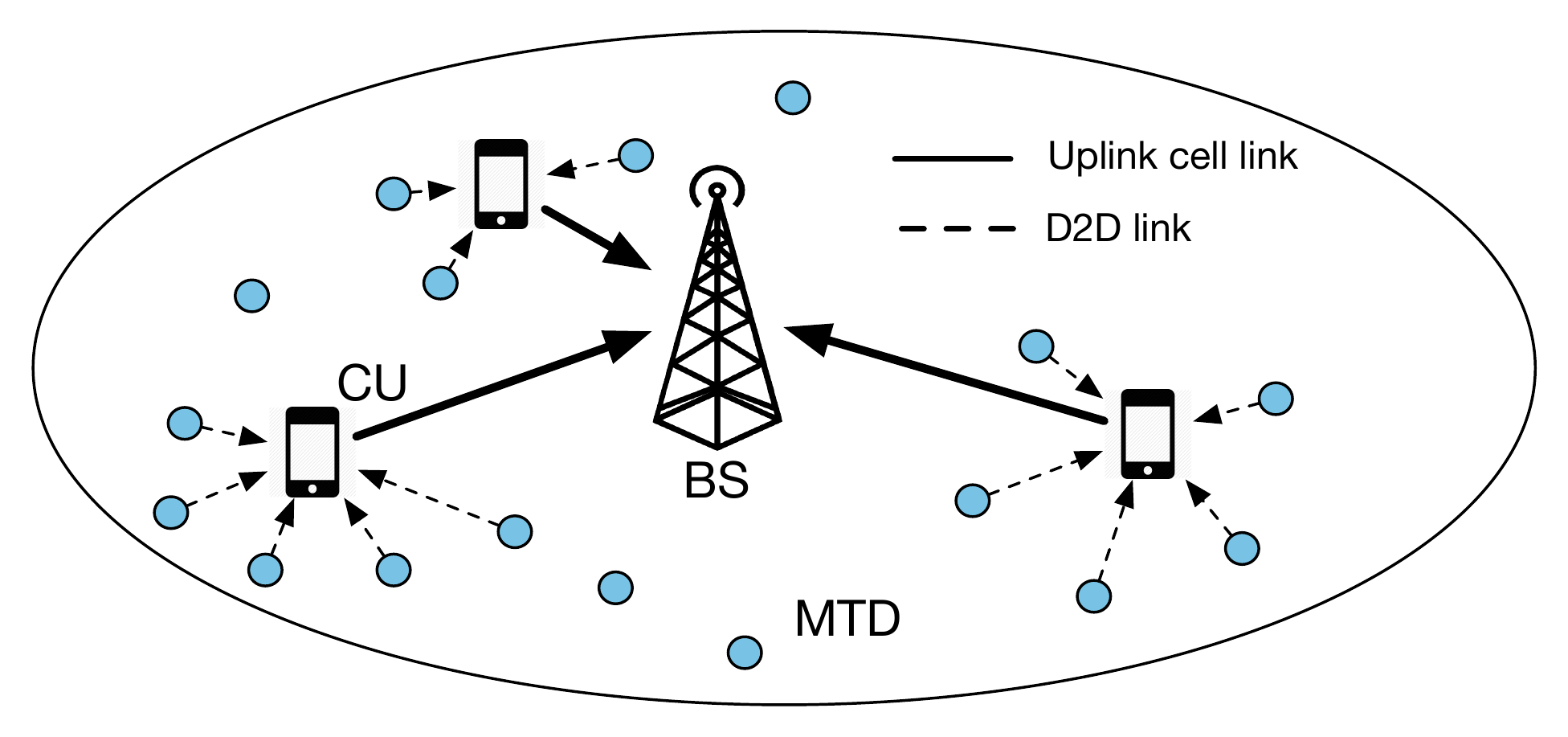}
\caption{MTDs and cellular users in a single cell network.}
\label{Fig0}
\end{center}
\end{figure}
Due to ubiquitous connectivity, reliable communications and high level of security, cellular networks represent a valid solution to accommodate the M2M traffic~\cite{UbiMassAccess}. Nevertheless, by being optimized for human-centric traffic patterns, network efficiency and scalability can be severely affected by the large number of active users, which can rapidly congest the access and core network due to signaling overhead.

An emerging communication paradigm in wireless cellular systems is Device-to-Device (D2D) communications, which is a form of Peer-to-Peer (P2P) communication~\cite{dohler}. This has inspired the idea to use cellular user devices as Machine Type  Gateways (MTG) and relay the data originated at the machines through D2D links. 
As an example, a network-assisted D2D technique to enable the cooperation between cellular users and MTDs is proposed in ~\cite{kiilerichunderlay}, where the underlay low-power D2D communications exploit Successive Interference Cancellation (SIC) to deliver the machine-type traffic to a device that subsequently acts as a relay. 

In literature, packet aggregation techniques have been proposed in diverse network architectures and systems. As an example, data aggregation and compression are extensively applied in clustered Wireless Sensor Networks (WSN), in order to reduce the energy consumption and increase the battery life time of the sensor nodes~\cite{Mhatre200445}. In the context of M2M, a random access scheme is investigated in~\cite{Cioffi}, where sensor devices and data collectors are randomly deployed within a cell, while  an approach to group a number of machines into a \emph{swarm} to alleviate the Radio Access Network (RAN) overload and reduce the number of connections between the devices and the BS is presented in~\cite{Chen20143}. 
This also allows to aggregate traffic packets originated from low-power Machine-Type Devices (MTDs) and dispatch them to a Base Station (BS), preventing the network access to overload and accommodate devices characterized by poor communication links~\cite{6133603}. 
However, relay-based schemes implicitly assume the presence of helper nodes, which not only increase the network operational expenditure, but also consume some of the radio resources to transmit the aggregated traffic to the BS, thus leading to the degradation of the network performance~\cite{5737888}.
%

Inspired by the paradigm of \emph{Trunked Radio Systems}~\cite{140487}, where limited radio resources are shared among a large set of users, we propose an access protocol that can potentially mitigate access overload and exploit the benefits of D2D. 
As shown in Fig.~\ref{Fig0}, packets generated by several MTDs are collected by a nearby Cellular User (CU) through D2D links, which is in charge of aggregating and delivering the traffic to a BS.
Our protocol consists of an access reservation phase, where the machines contend for access, and a Time Division Multiple Access (TDMA) scheme, where the time is divided in slots and each previously granted device is allocated a Time Slot (TS). After aggregating the packets and adding its own data, the mobile device transmits to the BS by adapting the power and the transmission rate to the channel conditions as well as the actual amount of data that needs to be sent. 
We show that there is a fundamental trade-off between latency and power required for the uplink transmission in an M2M scenario consisting of a large number of machines. 
We compare our technique with a traditional cellular access system, where the machines access directly to the BS, and we thereby demonstrate the power benefits of the trunking scheme.  

The remainder of this paper is organized as follows. Section~\ref{SysModel} describes the adopted system model. In Section~\ref{Analysis} a detailed analysis of the protocol is conducted and the numerical results are presented in Section~\ref{NumResults}. Conclusions are drawn in Section~\ref{Conclusions}.  

\section{System Model and Communication Framework} 
\label{SysModel}

\subsection{System Model}\label{subsec:sysModel}

\begin{figure}[t]
    \begin{center}
        \includegraphics[width=0.75\columnwidth]{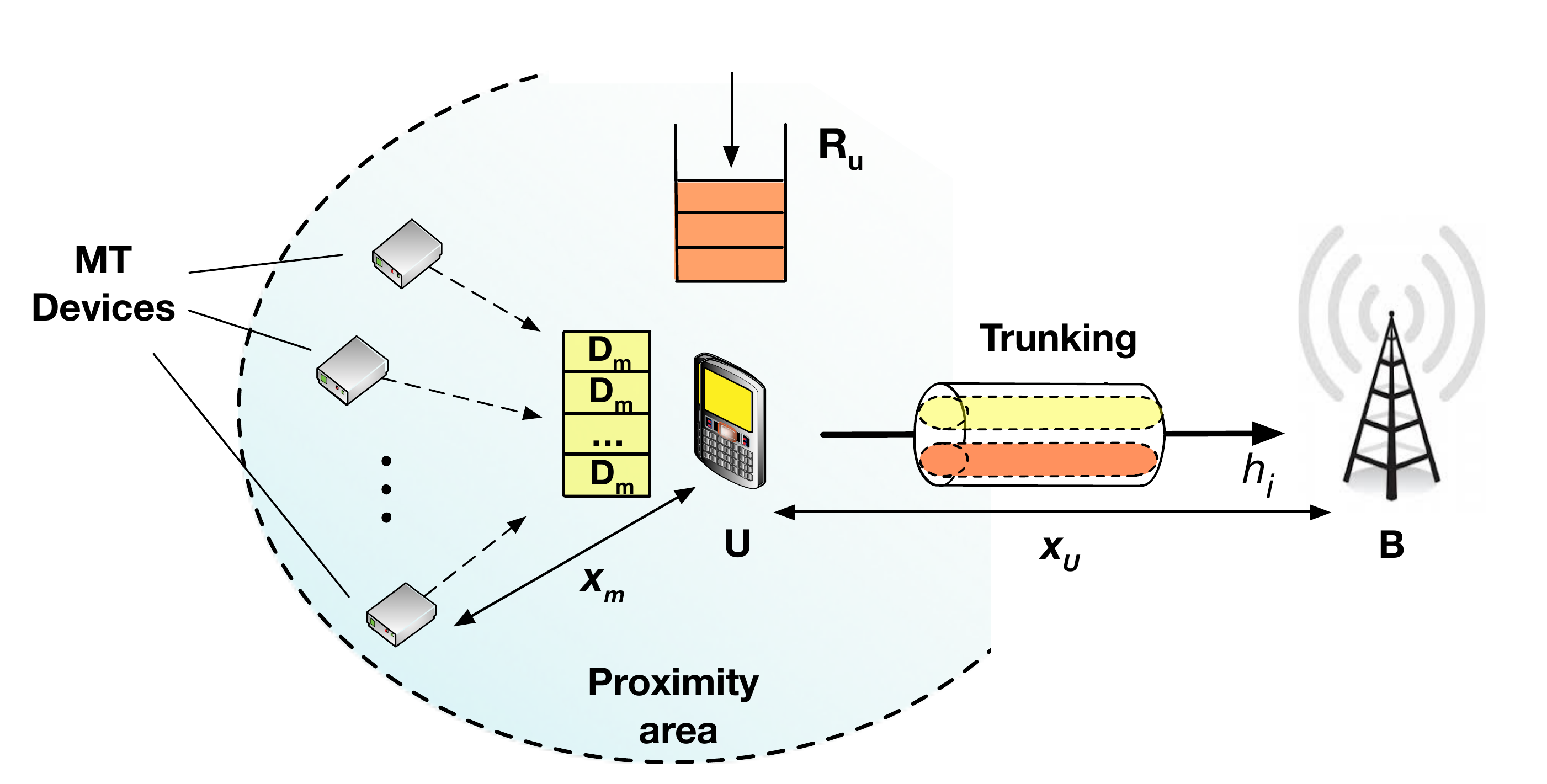}
        \caption{Illustration of the system model.}
        \label{Fig2}
    \end{center}
\end{figure}

Our system consists of a single cell in which a Base Station $B$ provides cellular access to a subscribed user $U$, and an M2M network composed by several low-power MTDs connected to $U$ via D2D communication. The time is divided in frames with duration $L \cdot T$, where $L$ is the fixed number of slots in the frame and $T$ is the slot duration. D2D communication takes place in a subset of time slots from a frame allocated to the user $U$, such that it can be characterized as overlay/in-band D2D~\cite{6805125}. As shown in Fig.~\ref{Fig2}, the MTDs are located in a proximity area around $U$, which corresponds to the maximum D2D coverage range, such that a D2D communication session can be established between $U$ and each nearby MTD.
For simplicity, we suppose that all the machines are at the same distance $x_m$ from $U$ and transmit at fixed power $P_m$, whereas the distance $U$-$B$ is fixed to $x_U$.
 The arrival process for machine-type data at the MTDs is modeled according to the Poisson distribution with arrival rate of $\lambda$ packets per second. We assume that each MTD can generate at most one packet of $D_m$ bits per frame, such that the packet arrivals to the population of MTDs are distributed at different MTDs.
 To evaluate the performance of the trunking scheme, we also suppose that the cellular uplink is constantly used by $U$ to transmit its own data at a constant bit rate of $R_u$. In the sequel, $U$ is assumed to be in charge of: $i)$ gathering the machine-type traffic received through $D2D$ links, $ii)$ aggregating the MTDs data with its own packets, and $iii)$ forwarding the aggregated traffic to $B$ via the cellular uplink. 
Furthermore, all the packets must be transmitted within a time interval equal to the frame duration $L \cdot T$, which represents the packet deadline.

All devices are equipped with a single antenna at both the transmitter and receiver side and $U$ has full and instantaneous channel state information (CSI) of the link towards $B$\footnote{Note that the absence of full CSI leads to the degradation of the uplink performance and consequent increase in the power consumption. The study of this kind of scenario is left for future work.}, allowing $U$ to perform uplink power control. We assume that all links are characterized by a block fading channel, where the channel state remains constant over the frame period and the fading realization follows a Rayleigh distribution with Probability Density Function (PDF), $f_h$, defined as,
\begin{equation}\label{RayleighPdf}
    f_{h}(u)= \frac{1}{\overline{h}}\exp\left(-\frac{u}{\overline{h}}\right),
\end{equation}
where $\overline{h}$\footnote{We assume $\overline{h} = 1$ in the remaining of the paper.} is the mean channel gain associated with the small scale fading. 
We define the Signal-to-Noise Ratio (SNR) from the $j^{th}$ transmitter, $\gamma_j$, as,
\begin{equation}
    \gamma_j = \frac{P_j h_j x_j^{-\alpha} K_D}{\sigma^2},
\end{equation}
where $P_j$ is the transmission power of the $j^{th}$ node, $x_j$ is the distance between the receiver and the $j^{th}$ node, $\alpha$ is the path loss exponent, $K_D$ is the path loss constant and $\sigma^2$ is the noise variance.
The list with all the mathematical symbols used in the paper is illustrated in Table~\ref{Tab_1}.

\begin{figure}[t]
    \begin{center}
        \includegraphics[width=0.8\columnwidth]{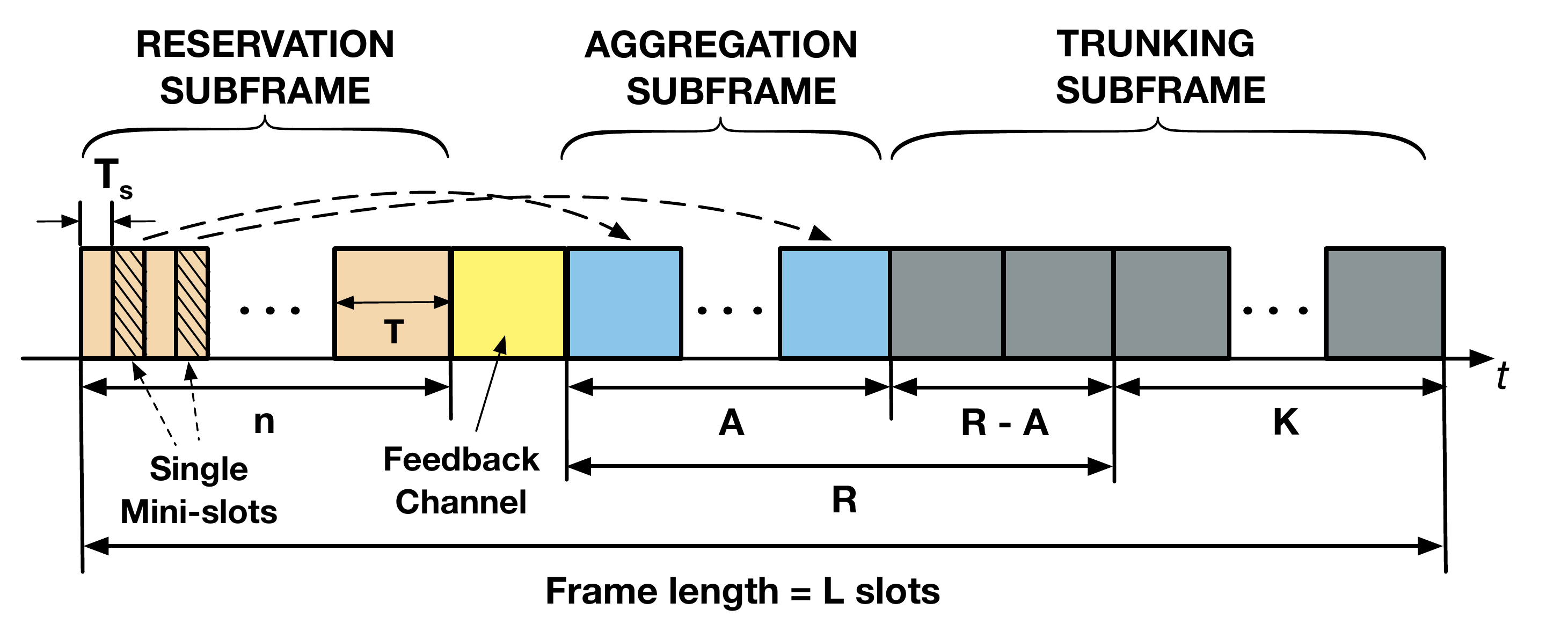}
        \caption{Frame structure of the proposed MAC protocol.}
        \label{Fig3}
    \end{center}
\end{figure}
\subsection{Proposed Communication Framework}\label{sec.proposedFramework}

Both D2D and uplink cellular communication take place within a TDMA frame of length $L$, as illustrated in Fig.~\ref{Fig3}. One-shot transmission scheme is employed, such that a packet that is not successfully received is dropped, i.e. not present in the next frame. The frame is divided into three orthogonal phases: reservation, aggregation and trunking, described in the sequel. 

\subsubsection{Reservation}

The first portion of the frame is dedicated to an access reservation phase that allows the MTDs to use random access and indicate their need for data transmission. A TS is divided into a 
fixed number of reservation mini-slots $R$ of duration $T_s$, such that the number of time slots allocated during this phase $n$ is:
\begin{equation} \label{value_of_n}
n = \lceil R \cdot \frac{T_s}{T} \rceil,
\end{equation}
where $R$ is the number of reservation mini-slots. 
The access reservation procedure uses~\emph{Framed Slotted ALOHA}~\cite{FramedALOHA}, where a MTD with a data packet to transmit, selects randomly and uniformly one of the available reservation mini-slots to transmit its reservation token. When multiple MTDs select to transmit in the same mini-slot, we assume that a destructive collision occurs, which leads to none of the reservation tokens being detected in that mini-slot. At the end of this phase, a mini-slot is declared as reserved only if a single MTD has selected it and the associated transmission is strong enough to be detected by the receiver. In other words, even if there is a single MTD that selects the mini-slot, the channel can be in outage such that the reservation token is not decoded and $U$ does not grant data access to the corresponding MTD. The next slot in the frame, as depicted in Figure~\ref{Fig3}, is reserved for the transmission of the necessary feedback to the MTDs through a channel with very robust coding, so that any errors occur with very low probability.
\begin{table}[tbp]
\renewcommand{\arraystretch}{1.1}
\footnotesize
\caption{Mathematical Notation}
\centering
\begin{tabular}{|ll|ll|} \hline 
    Sym      & Definition   & Sym      & Definition \\\hline \hline
    $\lambda$ & MTD arrival rate & $D_m$              &      MTD packet payload  \\ 
     $\lambda_f$ & Avg no of MTDs per frame   &   $D_u$              &      User packet payload  \\ 
     $\Lambda_i$ &   No of MTDs in frame $i$ & $h$              &      Channel power gain\\  
     $p_d$ &    Prob of successful D2D TX & $P_{U,i}$              &   $U$-$B$ req. TX power \\ 
     $P_{O}$ &   Outage probability in U-B & $T_s$              &     Mini-slot duration \\ 
    $L$ & Frame length   & $T$              &      Slot duration\\ 
    $K$ & No. of trunking slots  & $\mu$              &      TCI cutoff parameter  \\ 
    $R$           & No. of mini-slots  & $\Gamma_m$              &     $m$-$U$ SNR threshold   \\ 
    $A$           & No. of scheduled D2D slots & $R_{i,a}$              &      Aggregated data rate   \\ 
    $n$              &     No. of access reservation TSs & $W$	           &	    System bandwidth  \\   
    $P_m$ & MTD TX power TSs & $K_D$	   &	 Path loss factor      \\ 
    $x_j$ & Receiver-node $j$ distance & $\sigma^2$	   &	 Noise variance      \\ 
    $\gamma_j$ & Receiver-node $j$ SNR & $a$	   &	 No of succ. mini-slots      \\ 
     \hline          
    \end{tabular}
    \normalsize
    \label{Tab_1}
\end{table}

\subsubsection{Aggregation}

The aggregation phase, as depicted in Figure~\ref{Fig3}, is where the D2D communication between the MTDs that were granted access and the $U$ takes place. The number of slots given for data transmission to the MTDs is $A \leq R$, where $A$ corresponds to the number of successful reservations, excluding the ones that would lead to outage in the data transmission part. The system has $R$ preallocated slots for MTD data transmission, such that when $A$ is known, $R-A$ slots are returned back to $U$ for uplink transmission. Once all the scheduled machines have completed the D2D transmission, $U$ aggregates the machine packets with its own data.

\subsubsection{Trunking}

The last phase corresponds to the trunking phase, where the cellular uplink is used as a trunk to deliver both the received machine traffic and the user data to $B$. The device $U$ uses power control in the uplink, based on the Channel State Information at the Transmitter (CSIT) as well as the total amount of aggregated data to transmit, as described in Sec.~\ref{subsecTrunking}.
Finally, we note that the number of slots in the trunking phase is equal to $K + (R-A)$, where $K$ is the fixed number of reserved trunking slots, and $R-A$ is variable and depends on the number of slots not utilized in the aggregation phase. In conclusion, a single TDMA frame consists of a fixed number of slots $L$, divided as follows:
\begin{equation}
    L = n + R + K + 1.  
    \label{Frame_Length}
\end{equation}

\section{Analysis } \label{Analysis}

\subsection{Reservation}

A successful aggregation occurs when (a) the reservation slot has been selected by a single MTD and (b) the channel MTD-$U$ is sufficiently strong not to lead to outage. Given $R$ mini-slots, the probability to have $a$ successfully reserved mini-slots is 
\begin{equation}\label{eq:A}
    \text{Pr}\left( A = a | R \right) = \sum_{s=1}^R \text{Pr}\left( A = a | s\right) \text{Pr}\left( S = s | R\right),
\end{equation}
where $\text{Pr}\left( S = s | R\right)$ is the probability of having $s$ single mini-slots:
\begin{equation}
    \text{Pr}(S = s | R) = \sum_{m=0}^{\infty} \text{Pr}(S=s | m, R)  \text{ Pr}(\Lambda_i = m),
    \label{probS_s}
\end{equation}
where $\text{Pr}(\Lambda_i = m)$ is the probability of having $m$ MTDs arrivals in the $i^{th}$ frame.   Let $\lambda$ be the MTDs arrival rate per second and $\lambda_f = E[\Lambda_i] = \lambda L T$ be the average number of MTDs transmitting at the beginning of $i^{th}$ frame. Then $\Lambda_i$, assuming stationarity and independence from the channel states, is given by a Poisson distributed process with arrival rate $\lambda_f$ and associated probability mass function,
\begin{equation}
    \text{Pr}(\Lambda_i=m) = \frac{ \lambda_f^m e^{- \lambda_f}}{m!}, m \geq 0.
    \label{pmf}
\end{equation}
In (\ref{eq:A}), the probability $ \text{Pr}\left( A = a | s\right)$  to have $a$ successful reservations given $s$ single mini-slots is 
\begin{equation}
    \text{Pr}\left( A = a | s\right) = \binom{s}{a} p_d (1-p_d)^{s-a},
\end{equation}
where $p_d$ is the probability that the reservation token is successfully decoded. Assuming $\Gamma_M$ as the minimum SNR required for the transmission from a MTD to be decodable, then,
\begin{equation}\label{eq:pd}
    p_d = 1 - \text{Pr}(\gamma_{m} < \Gamma_{m}) = e^{-\Gamma_{m}\frac{\sigma^2}{P_m \bar{h}_i x_m^{-\alpha} K_D}}.
\end{equation}

To compute $\text{Pr}(S=s | m,R)$, we follow the approach described in~\cite{vogt2002efficient}. The process of distributing MTDs into mini-slots can be modeled as a bins and balls problem, with distinguishable bins and balls. In this context, the bins and the balls are represented by the mini-slots and the MTDs, respectively, and the number of single mini-slots corresponds to the number of bins with occupancy number equal to $1$. By taking into account the results presented in~\cite{vogt2002efficient}, the probability of having exactly $s$ out of $R$ mini-slots given $m$ contending MTDs is:
\begin{equation}
    \text{Pr}(S=s | m, R) = \frac{\binom{R}{s} \prod_{k=0}^{s-1} (m-k) G(R-s,m-s)} {R^m},
    \label{BallsAndBins}
\end{equation}
where\footnote{We define $\binom{n}{k}=0$ for $n<k$.}
\begin{equation}
    G(u,v) = u^v + \sum_{t=1}^{v} (-1)^t \prod_j^{t-1}[(v-j)(u-j)](u-t)^{v-t}\frac{1}{t!}.
    \label{G}
\end{equation}

To ease the computation of~\eqref{probS_s}, we provide the following approximation. Assuming that the probability of a MTD choosing a random mini-slot out of $R$ mini-slots is $1/R$, leading to the mean value of the number of contending MTDs per mini-slot to be equal to $\lambda_f/R$. Thus, the probability of a single MTD selecting a mini-slot can be approximated as,
\begin{equation}
p_{s} \approx \frac{ \lambda_f}{R} e^{- \frac{\lambda_f}{R} }.
\label{probsingle}
\end{equation}
Now, by assuming the amount of MTDs in each mini-slot is independent, then the probability that $s$ out of $R$ mini-slots are single is,
\begin{equation}
    \text{Pr}(S=s | R) \approx \binom{R}{s} p_{s}^s (1-p_{s})^{R-s},
    \label{approxprobS_s}
\end{equation}
where $(1-p_{s})^{R-s}$ is the probability that none of the remaining $R-s$ mini-slots is single, and $\binom{R}{s}$ is the number of ways this slot can be selected. This approximation becomes tighter with the increase of the number of arrivals in the frame.

\subsection{Aggregation}

After the access reservation phase, each of the $a$ accepted MTDs are allowed to transmit via the D2D link to $U$, through their own dedicated data slot. 
As a result, the aggregated data rate in the $i$-th frame $R_i$ becomes,
\begin{equation}
    R_{i,a} = \frac{D_u + a D_m}{T(K + R - a)},
\end{equation}
where $D_m$ is the MTDs packet payload, while $D_u$ is the user payload, generated every frame time period $LT$ with rate $R_u$ as $D_u = L T R_u$.
\subsection{Trunking}\label{subsecTrunking}

As mentioned in Sec.~\ref{sec.proposedFramework}, we consider a cellular uplink with adaptive rate allocation, where the user transmitter is able to dynamically adjust the power depending on the data rate and on the channel conditions.
Assuming AWGN channel and capacity-achieving codes, the data rate in the $i$-th frame $R_i$ is related to the transmitting power $P_{U,i}$ by the Shannon's capacity equation:
\begin{equation}
    R_{i,a} = W \log_2 (1 + \frac{P_{U,i} h_i x_j^{-\alpha} K_D}{\sigma^2}),
    \label{dataRate}
\end{equation}
from which $P_{U,i}$ is obtained as,
\begin{equation}
    P_{U,i} = \left(2^{\frac{R_{i,a}}{W}}-1\right)\frac{\sigma^2}{h_i x_j^{-\alpha} K_D},
\end{equation}
which is the transmit power needed to sustain the data rate $R_i$ over the channel during the $i$-th frame. 
To compensate for the variation of the channel between $U$ and $B$, we consider a Truncated Channel Inversion (TCI) policy~\cite{775366}, where the channel fading is inverted only if the fade depth is above a given cutoff value $\mu$. Therefore, the average transmit power required to sustain the data rate $R_{i,a}$ is derived as follows: 
\begin{align}
    \mathbb{E}[P_{U,i}|R_{i,a}]&=\int\limits_\mu^\infty P_{U,i} ~f_{h}(x) dx \\ \nonumber
                       &=  \left(2^{\frac{R_{i,a}}{W}}-1\right)\frac{\sigma^2}{\bar{h}_i x_U^{-\alpha} K_D} E_1\left(\frac{\mu}{\bar{h}}\right),
    \label{E_TCI}
\end{align}
 where $\mu$ is the cutoff parameter representing the minimum value of channel fade depth that can be compensated and $E_1(\cdot)$ is the exponential integral function. Accordingly, the related outage probability is given by:
\begin{equation}\label{eq:Pout}
    P_{O} = \text{Pr}(h < \mu) = \int\limits_0^\mu f_{h}(x) dx = 1 - \exp\left(-\frac{\mu}{\bar{h}}\right).
\end{equation}
Note that, to guarantee a certain outage probability, we can conveniently select the value of $\mu$, as follows,
\begin{equation}
\mu =  -\bar{h}\log(1-P_{O}).
\end{equation}

\section{Numerical Results} \label{NumResults}
\subsection{Performance Metrics}
We now provide the analytical expressions for the performance metrics which are examined in the next section. As first metric, we consider the expected number of MTDs served per second, defined as:
\begin{equation}
    \mathbb{E}[N] = \frac{1}{L \cdot T} \sum_{a=0}^{R} \text{Pr}\left( A = a | R \right),
\label{AvgNbMTCDsServed}
\end{equation}
where $p_d$ denotes the probability of the data slot being decodable by the receiver and is obtained from~\eqref{eq:pd}, while $\text{Pr}\left( A = a | R \right)$ is the probability of having successfully reserved $a$ mini-slots given $R$ mini-slots, obtained from~\eqref{eq:A}.

Next, the average transmit power per served machine $\mathbb{E}[P_m]$ is considered, which consists of the power needed to reserve a time slot $P_{res}$ plus the power to transmit the data packet $P_{agg}$ and the power to send the machine packet through the trunk uplink channel $\mathbb{E}[P_{tr}]$:
\begin{equation}
    \mathbb{E}[P_m] =  P_{res} + P_{agg} + \mathbb{E}[P_{tr}] = 2 P_{m} + \mathbb{E}[P_{tr}],
\label{AvgNbMTCDsServed}
\end{equation}
where $P_{res} = P_{agg} = P_m$. Depending on whether the MTDs are connected to $U$ or $B$,  we assume $P_m = P_{m,U}$ or $P_m = P_{m,B}$ respectively, while $\mathbb{E}[P_{tr}]$ is given by,
\begin{equation}
    \mathbb{E}[P_{tr}] = \sum_{a=0}^{R}\mathbb{E}[P_{U,i}|R_{i,a}] ~ \text{Pr}\left( A = a | R \right).
\label{AvgTXPwrTrunking}
\end{equation}

Finally, we define the probability of a user being served as the probability of being the only one to select a given mini-slot, given $m$ arrivals, and of the reservation token being detected in the reservation phase and the trunking link not being in outage:
\begin{equation}
    P_S = (1 - P_{O}) \sum_{m=0}^{\infty} p_{d} \left( 1 - \frac{1}{R} \right)^{m -1} \text{Pr}(\Lambda_i=m) .   
\end{equation}
\begin{table}[tbp]
\renewcommand{\arraystretch}{1.1}
\footnotesize
\caption{Simulation Parameters}
\centering
\begin{tabular}{llllll} \hline
    Par     & Value & Par      & Value   & Par      & Value \\ 
    \hline \hline

    $P_{O}$ & $0.01$ & $T_s$              &      $0.1$ ms  & $\sigma^2$	           &	    $-97$ dBm  \\ 
    $R_u$              &     $100$ Kbps & $W$	           &	    $180$ KHz & $K_D$ 		&  $-30$ dB \\  
    $D_m$              &     $100$ bits & $\Gamma_m$	           &	    $-3$ dB  & $x_{m} $	           &	    $10$ m  \\  
    $T$              &      $1$ ms & $\overline{h}$	           &	    1  & $x_{U} $ 		&  $200$ m \\ 
        $\alpha$              &      $3$  & $P_{m,B}$	 &  $18$ dBm & $P_{m,U} $ &  $-20$ dBm \\ 
     \hline
    \end{tabular}
    \normalsize
    \label{Tab}
\end{table}

\subsection{Simulation Results}\label{SimResults}
To numerically analyze the performance of our protocol, we developed a MATLAB-based simulator, implementing the functionalities described in Sec.~\ref{subsec:sysModel} and assuming the parameters shown in Table~\ref{Tab}. Simulations are based on a Monte Carlo approach, where each point corresponds to the average value of $10^5$ iterations.
We first examine the performance metrics by increasing $\lambda$, in order to observe the impact of different network access volumes on the system. 
Furthermore, we analyze the average power spent to serve a single machine and compare our approach with a typical cellular system, where the distance between MTDs and B is fixed to $x_{U}$ and all the machines communicate with $B$, using a fixed power level $P_m = P_{m,B}$ and without relying on the trunk link.
It is also worth pointing out that in our system the latency is only given by the frame length $L$, which depends on the parameters $R$ and $K$, as shown in (\ref{Frame_Length}).

\begin{figure}[tbp]
\centering
   \includegraphics[width=0.96\columnwidth]{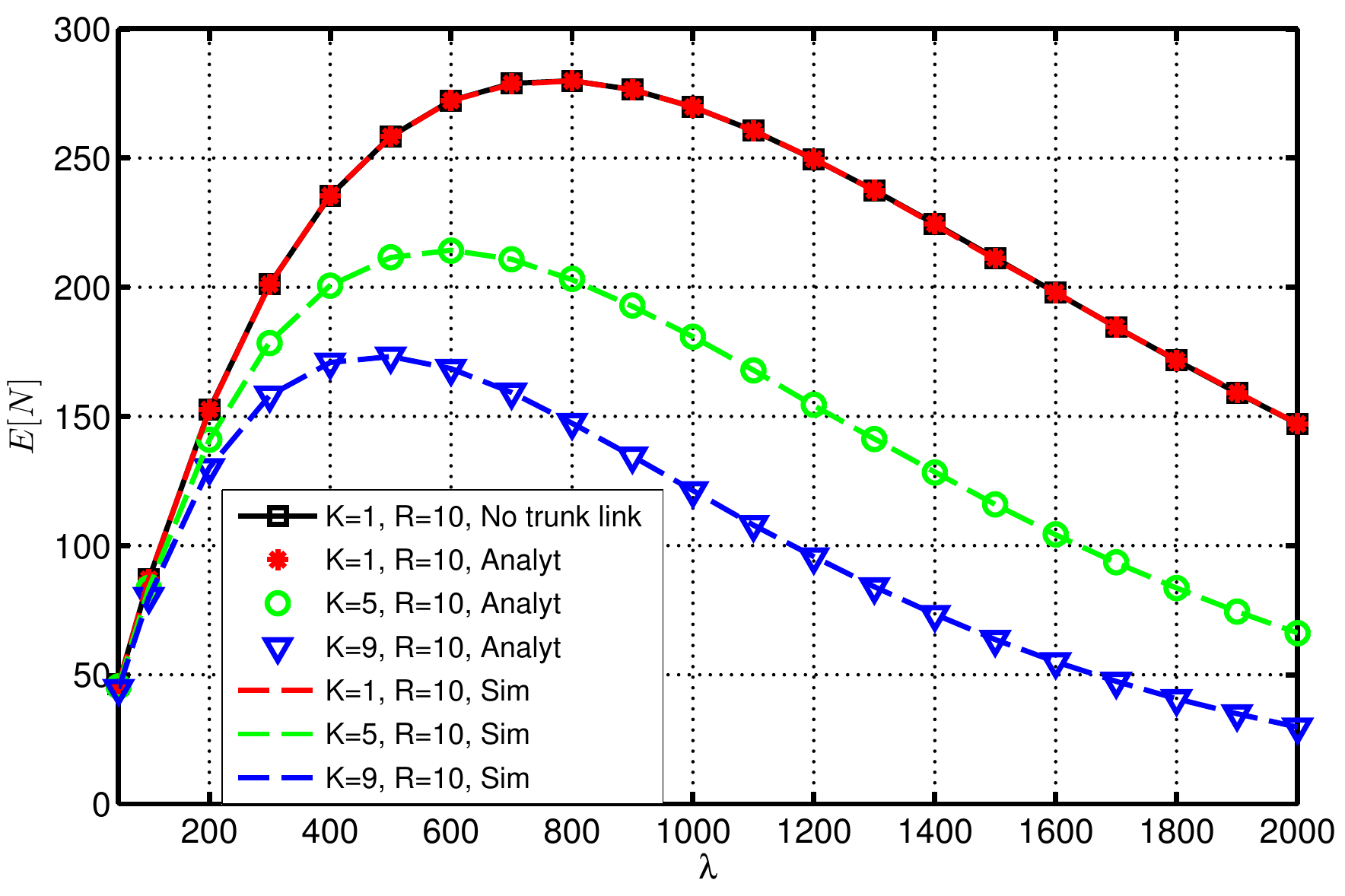} 
 \caption{Average number of served MTDs for different $K$ and $R$.}
\label{fig5}
\end{figure}

Fig.~\ref{fig5} shows the average number of machines served $\mathbb{E}[N]$ as a function of $\lambda$. 
Taking as reference the curve with parameters $K=1$ and $R=10$, we can observe that the number of selected single mini-slots increases and reaches its maximum value at $\lambda \approx 800$. Then, the impact of the collisions becomes evident and $\mathbb{E}[N]$ decreases. 
Moreover, choosing $R=10$ corresponds to assign $n=1$ slot to the access subframe, which increases the protocol overhead as well as the system latency.
We note that a higher $K$ significantly affects the system performance, since a longer frame results in a larger number of contending machines due to the aforementioned Poisson arrivals assumption. 
This, in turn, leads to an increased number of collisions in the reservation subframe and a consequent decrease of $\mathbb{E}[N]$. 
In addition, it is worth noticing that the expected value of the total power transmitted over the $U$-$B$ link is directly related  to $\mathbb{E}[N]$, as the more machine packets are aggregated, the higher power is required to sustain the resulting data rate. 
Therefore, the maximum required transmit power can be obtained by considering values of $\lambda$ which maximize $\mathbb{E}[N]$, according to $R$ and $K$.
\begin{figure}[tbp]
\centering
   \includegraphics[width=0.96\columnwidth]{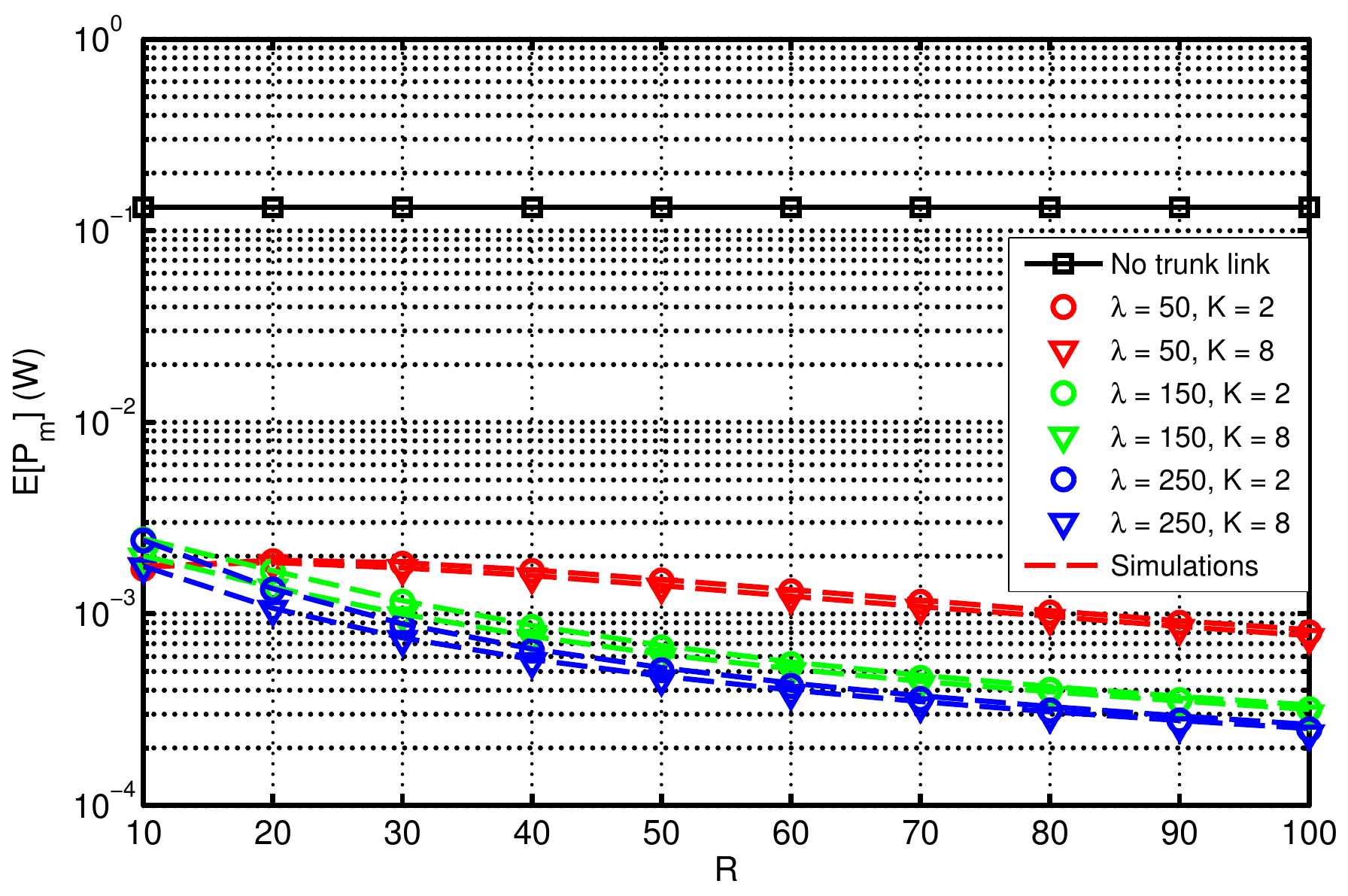} 
\caption{Average transmit power per served MTD for different $\lambda$ and $K$.}
\label{fig8}
\end{figure} 

The trade-off between latency and transmit power is illustrated in Fig.~\ref{fig8}.
First of all, we notice a significant difference in terms of consumed power between our scheme and a system without trunk link, due to the fact that in the latter case the MTDs transmit at fixed power and no dynamic power allocation is allowed.
Secondly, if we consider the trunk link approach with $\lambda = 250$ and choose $R=20$, the power consumption is approximately reduced by $50\%$ with respect to the case of $R = 10$. In other words, by increasing $R$, not only the number of served machines increases, but also the number of available trunking slots grows, since we assumed the system able to reuse the portion of the frame not occupied by the D2D transmissions, i.e., $R-A$.
In contrast, the latency of the system becomes higher, as we need to assign more slots to the reservation subframe, and this also implies a significant increment in the overhead.
\begin{figure}[tbp]
\centering
   \includegraphics[width=0.96\columnwidth]{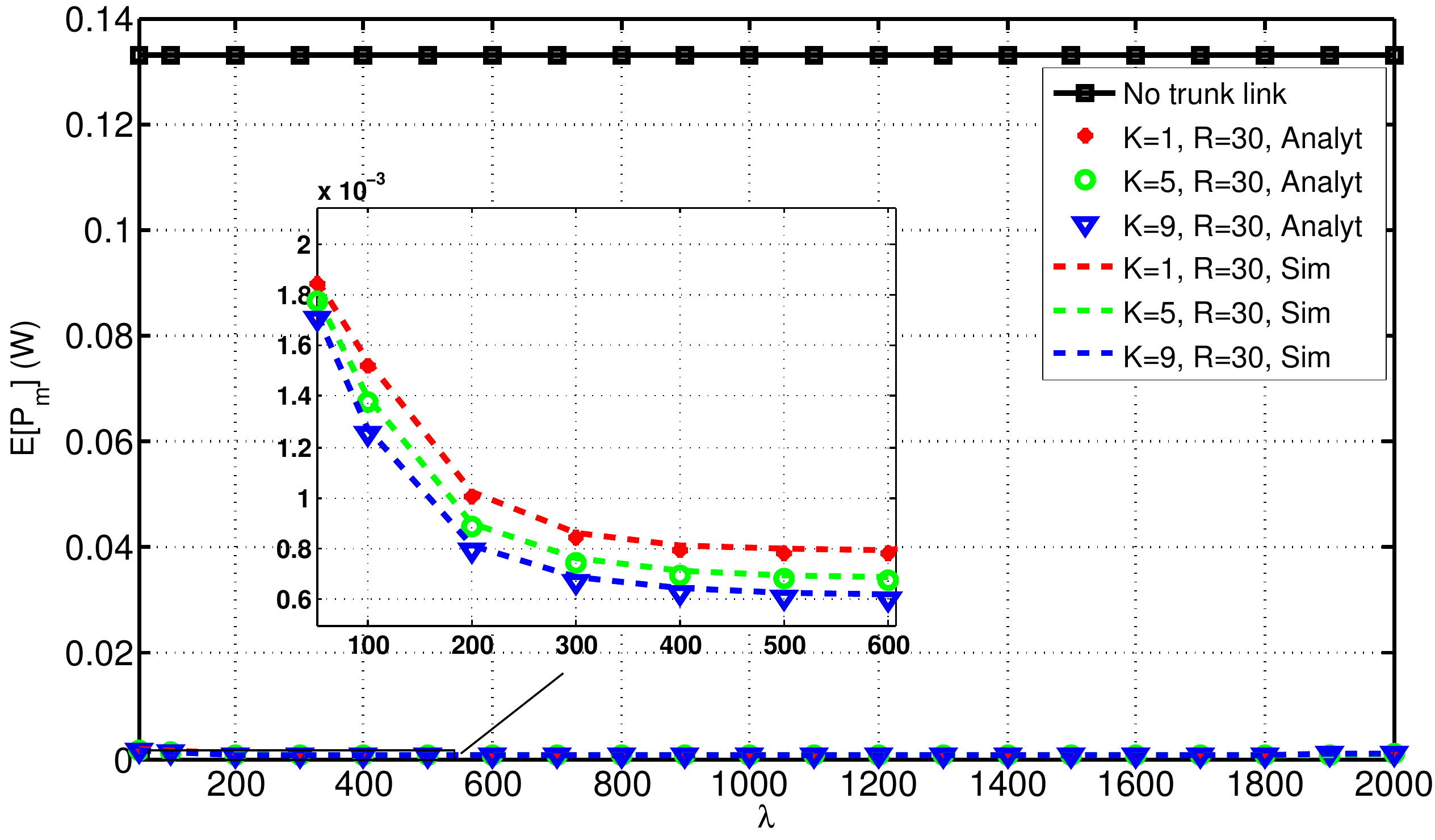} 
\caption{Average transmit power per served MTD for different $K$ and $R$.}
\label{fig6}
\end{figure} 
\begin{figure}[tbp]
\centering
   \includegraphics[width=0.96\columnwidth]{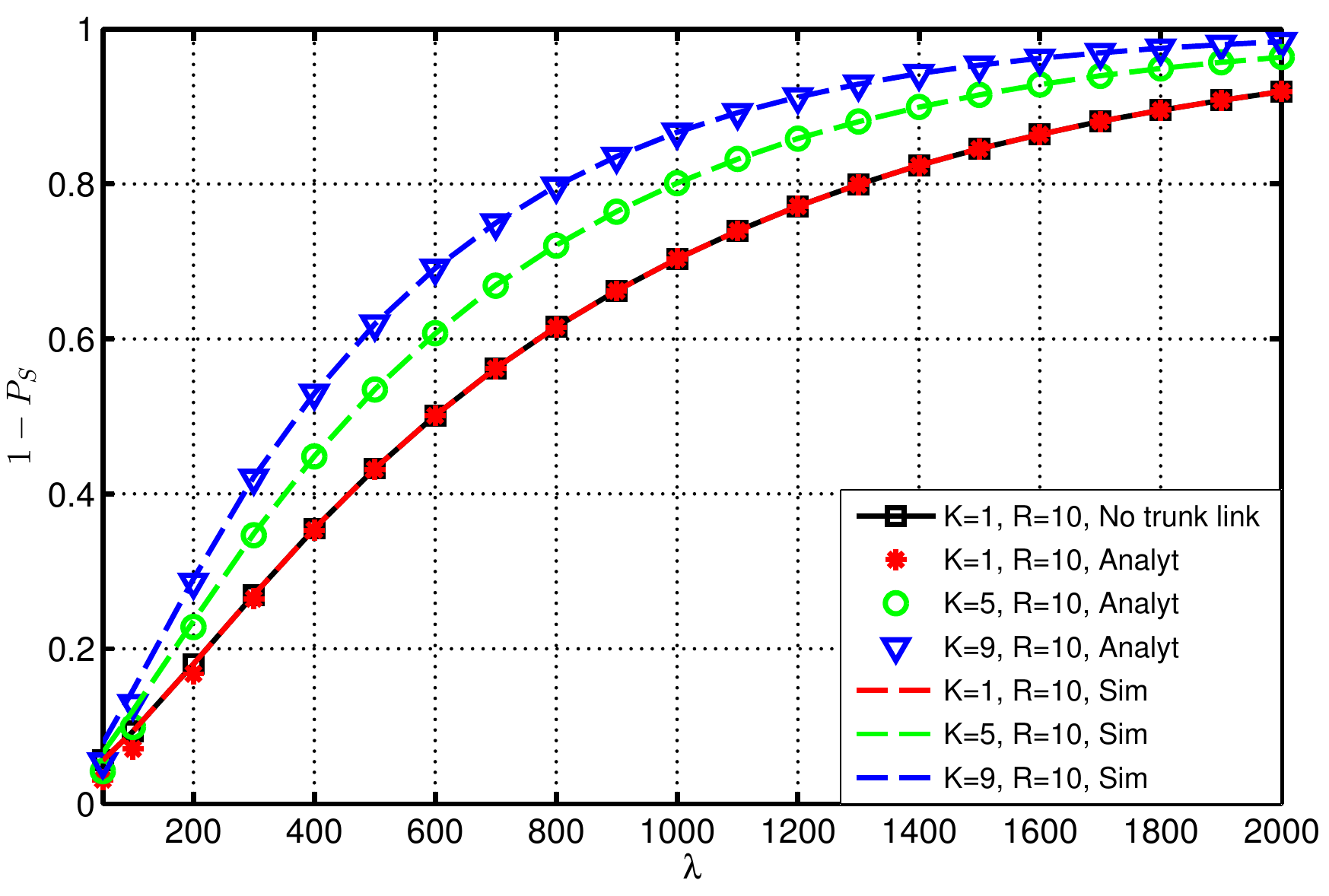} 
\caption{Machine outage probability for different $K$ and $R$.}
\label{fig7}
\end{figure} 

Fig.~\ref{fig6} shows the mean transmit power spent per MTD as a function of $\lambda$. 
As previously noticed, the power spent to transmit a single machine packet in the case of trunk link not available is fixed and does not depend on the number of contending MTDs. On the other hand, if the machines utilize the trunk link provided by $U$, $\mathbb{E}[P_{tr}]$ decreases as $\lambda$ becomes higher and turns out to be considerably lower than the case of trunking not enabled. 
To evaluate the impact of different number of trunk slots, we also included a zoomed in section illustrating $\mathbb{E}[P_{tr}]$ in the case of low load access conditions. More specifically, we can notice how the power decreases as $\lambda$ increases, since a higher number of machines is allowed to transmit, and the minimum value corresponds to the value of $\lambda$ maximizing $\mathbb{E}[N]$.

Finally, the outage probability of a machine device requiring access is shown in Fig.~\ref{fig7}. In case of low access load (i.e., $\lambda < 100$), the outage probability is below $0.2$ and a higher $K$ does not affect significantly the system performance. 
By increasing $\lambda$, the outage probability becomes higher due to the large number of access attempts causing collisions in the reservation phase. As expected, for higher access load (i.e., $\lambda > 400$), the outage probability is clearly higher if we assume $K>1$ and this also implies a lower number of machines served with respect to the case of a more energy-consuming system characterized by $K=1$. 

\section{Conclusions}\label{Conclusions}
We proposed a new solution to alleviate the impact of the massive number of M2M devices on a cellular system by exploiting D2D communications. Our TDMA-based MAC scheme guarantees the simultaneous delivery of packets  generated by the MTDs and by a user device, without involving any relay or helper node. The analytical and simulation results show that there is a fundamental trade-off between the latency and the transmit power in the case of Poisson-distributed machine arrivals. By increasing the frame length, we can achieve a remarkable reduction of the transmit power, even though the system latency increases and a low delay packet requirement cannot be fulfilled. Moreover, our scheme can significantly reduce the average amount of power spent to sustain a MTD in comparison with a system without trunk link, guaranteeing the same user throughput.

\bibliographystyle{IEEEtran}
\bibliography{IEEEabrv,Biblio}

\begin{thebibliography}{10}
\providecommand{\url}[1]{#1}
\csname url@samestyle\endcsname
\providecommand{\newblock}{\relax}
\providecommand{\bibinfo}[2]{#2}
\providecommand{\BIBentrySTDinterwordspacing}{\spaceskip=0pt\relax}
\providecommand{\BIBentryALTinterwordstretchfactor}{4}
\providecommand{\BIBentryALTinterwordspacing}{\spaceskip=\fontdimen2\font plus
\BIBentryALTinterwordstretchfactor\fontdimen3\font minus
  \fontdimen4\font\relax}
\providecommand{\BIBforeignlanguage}[2]{{%
\expandafter\ifx\csname l@#1\endcsname\relax
\typeout{** WARNING: IEEEtran.bst: No hyphenation pattern has been}%
\typeout{** loaded for the language `#1'. Using the pattern for}%
\typeout{** the default language instead.}%
\else
\language=\csname l@#1\endcsname
\fi
#2}}
\providecommand{\BIBdecl}{\relax}
\BIBdecl

\bibitem{6163599}
T.~Taleb and A.~Kunz, ``{Machine type communications in 3GPP networks:
  potential, challenges, and solutions},'' \emph{Communications Magazine,
  IEEE}, vol.~50, no.~3, pp. 178--184, March 2012.

\bibitem{4736537}
``{Health Informatics - PoC Medical Device Communication - Part 00101:
  Guide--Guidelines for the Use of RF Wireless Technology},'' \emph{IEEE Std
  11073-00101-2008}, pp. 1--99, Dec 2008.

\bibitem{UbiMassAccess}
S.-Y. Lien, K.-C. Chen, and Y.~Lin, ``{Toward ubiquitous massive accesses in
  3GPP machine-to-machine communications},'' \emph{Communications Magazine,
  IEEE}, vol.~49, no.~4, pp. 66--74, 2011.

\bibitem{dohler}
K.~Zheng, F.~Hu, W.~Wang, W.~Xiang, and M.~Dohler, ``{Radio resource allocation
  in LTE-advanced cellular networks with M2M communications},''
  \emph{Communications Magazine, IEEE}, vol.~50, no.~7, pp. 184--192, 2012.

\bibitem{kiilerichunderlay}
N.~{K. Pratas} and P.~Popovski, ``{Underlay of Low-Rate Machine-Type D2D Links
  on Downlink Cellular Links},'' in \emph{ICC, 2014 IEEE}, June 2014.

\bibitem{Mhatre200445}
``Design guidelines for wireless sensor networks: communication, clustering and
  aggregation,'' \emph{Ad Hoc Networks}, vol.~2, no.~1, pp. 45 -- 63, 2004.

\bibitem{Cioffi}
T.~Kwon and J.~Cioffi, ``{Random Deployment of Data Collectors for Serving
  Randomly-Located Sensors},'' \emph{Wireless Communications, IEEE Transactions
  on}, vol.~12, no.~6, pp. 2556--2565, June 2013.

\bibitem{Chen20143}
K.-C. Chen and S.-Y. Lien, ``{Machine-to-machine communications: Technologies
  and challenges},'' \emph{Ad Hoc Networks}, vol.~18, no.~0, pp. 3 -- 23, 2014.

\bibitem{6133603}
S.~Andreev, O.~Galinina, and Y.~Koucheryavy, ``{Energy-Efficient Client Relay
  Scheme for Machine-to-Machine Communication},'' in \emph{GLOBECOM, 2011
  IEEE}, Dec 2011, pp. 1--5.

\bibitem{5737888}
V.~Pourahmadi, S.~Fashandi, A.~Saleh, and A.~Khandani, ``Relay placement in
  wireless networks: A study of the underlying tradeoffs,'' \emph{Wireless
  Communications, IEEE Transactions on}, vol.~10, no.~5, pp. 1383--1388, May
  2011.

\bibitem{140487}
A.~Chrapkowski and G.~Grube, ``{Mobile trunked radio system design and
  simulation},'' in \emph{Vehicular Technology Conference, 1991. Gateway to the
  Future Technology in Motion., 41st IEEE}, May 1991, pp. 245--250.

\bibitem{6805125}
A.~Asadi, Q.~Wang, and V.~Mancuso, ``{A Survey on Device-to-Device
  Communication in Cellular Networks},'' \emph{Communications Surveys
  Tutorials, IEEE}, vol.~PP, no.~99, pp. 1--1, 2014.

\bibitem{FramedALOHA}
J.~Wieselthier, A.~Ephremides, and L.~Michaels, ``{An exact analysis and
  performance evaluation of framed ALOHA with capture},'' \emph{Communications,
  IEEE Transactions on}, vol.~37, no.~2, pp. 125--137, Feb 1989.

\bibitem{vogt2002efficient}
H.~Vogt, ``{Efficient object identification with passive RFID tags},'' in
  \emph{Pervasive Computing}.\hskip 1em plus 0.5em minus 0.4em\relax Springer,
  2002, pp. 98--113.

\bibitem{775366}
M.-S. Alouini and A.~Goldsmith, ``{Capacity of Rayleigh fading channels under
  different adaptive transmission and diversity-combining techniques},''
  \emph{Vehicular Technology, IEEE Transactions on}, vol.~48, no.~4, pp.
  1165--1181, Jul 1999.

\end{thebibliography}

\end{document}